\documentclass[prl,superscriptaddress,reprint]{revtex4-1}

\usepackage{amsmath}
\usepackage{graphicx}
\usepackage{bm}
\usepackage{MnSymbol}
\usepackage{eufrak}
\usepackage{amsfonts} 

\usepackage[usenames,dvipsnames,x11names]{xcolor}
\usepackage[normalem]{ulem}
\usepackage{tikz}
\usetikzlibrary{arrows}

\newcommand{\be}{\begin{equation}}
\newcommand{\ee}{\end{equation}}
\newcommand{\bea}{\begin{eqnarray}}
\newcommand{\eea}{\end{eqnarray}}

\newcommand{\braket}[1]{\left\langle   #1  \right\rangle}

\newcommand{\w}{\omega}

\begin{document}

\title{Electron shuttle as an autonomous single-electron source}

\author{Christopher W. W\"achtler}
\affiliation{Max-Planck-Institut für Physik komplexer Systeme, Nöthnitzer Str. 38, 01187 Dresden, Germany}
\author{Javier Cerrillo}
\address{\'Area de F\'isica Aplicada, Universidad Polit\'ecnica de Cartagena, E-30202 Cartagena, Spain}

\begin{abstract}
The majority of experimental realizations of single-electron sources rely on the periodic manipulation of the tunnel junctions through their gate voltages, and thus require a high level of control over the system. To circumvent the necessity of external driving, we utilize the self-oscillatory behavior of the electron shuttle. By means of waiting time distributions, which had not been applied to this autonomous system before, we extensively assess the  performance of the shuttle as a single-electron source. We unveil a smooth transition between three regimes, whereas previous studies at the same mean field level of description only predict a sharp bifurcation based on the time-averaged electron current. Over the parameter range of already existing experimental realizations the electron shuttle can perform as a single-electron source,  albeit with moderate precision. We propose an alternative design of the position-dependent tunneling rates, which significantly decreases the relative error of charge transmission, and thus paves the way for the realization of autonomous single-electron sources.
\end{abstract}

\maketitle

\section{Introduction}

The ability to fabricate devices capable of controlling single-charge transport in a highly precise manner has received considerable attention in the past years due to their applications in the field of quantum metrology \cite{zimmerman2003electrical, flowers2004route, keller2008current,pekola2013single}. In this context, electron pumps and turnstiles controlled via periodic modulation of gate voltages represent a common route to achieve single-electron sources (SES) \cite{blumenthal2007gigahertz, kaestner2008single, ono2003electron, pekola2008hybrid, giblin2012towards, fricke2014self, rossi2014accurate, yamahata2014gigahertz, brun2016practical, kaestner2015non, stein2015validation}. This limits the frequency of the devices due to the electronics involved and requires high accuracy of the drive. Furthermore, such conventional SES involving external driving become more deterministic as the coupling to the leads increases \cite{AlbertEtAlPRL2011}. However, when modeling the dynamics, a large coupling to the leads is problematic since that is a regime where typical rate equation descriptions fail and more sophisticated approaches are required, which may provide conflicting predictions \cite{RestrepoPRB2019}. 

\begin{figure}
\begin{center}
\includegraphics[width=\columnwidth]{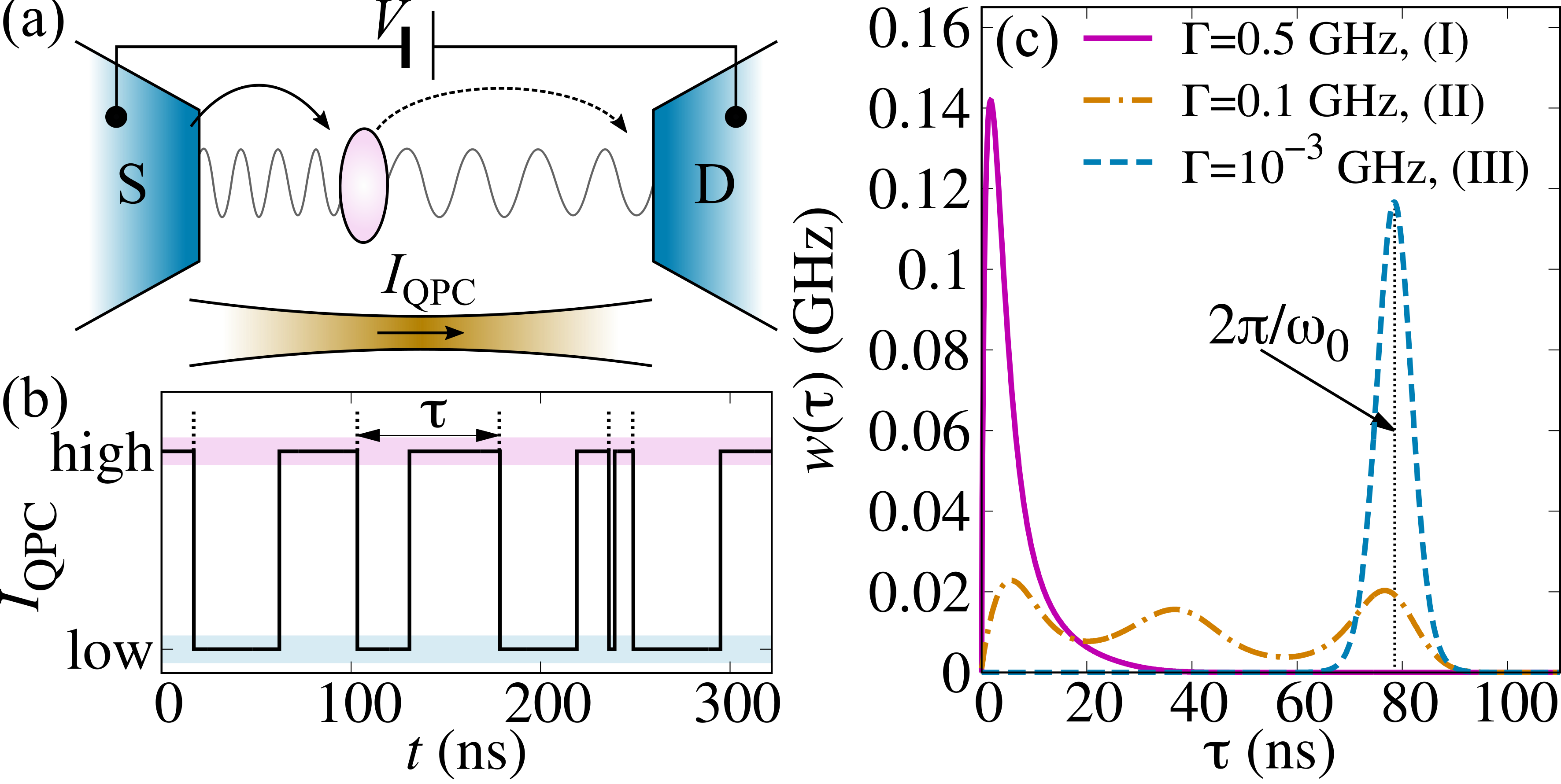}
\end{center}
\caption{(a) The electron shuttle: A quantum dot (pink) oscillates autonomously between two leads, source (S) and drain (D), to which it is tunnel coupled (black arrows). A nearby quantum point contact (QPC) can be used to monitor the shuttle occupation. (b) Typical time trace of the current through the QPC containing the waiting times $\tau$ between two consecutive tunneling events. (c) The waiting time distributions constructed from the time trace for three different values of the bare tunneling rate $\Gamma$. The oscillation period of the shuttle $2\pi/\omega_0$ is indicated, and we choose $V=70$~mV.}
\label{fig:Model}
\end{figure}

We investigate a different realization of SES by employing the phenomenon of self-sustained oscillations, where a system maintains a periodic motion in an autonomous fashion, thereby circumventing the necessity to externally drive the system. Self-sustained oscillations routinely arise in the context of electron shuttling experiments \cite{ParkEtAlNature2000, ErbeEtAlPRL2001, 
ScheibleBlickAPL2004, MoskalenkoEtAlPRB2009, MoskalenkoEtAlNanotechnology2009, KimNJP2010, KimPRL2010, 
KimEtAlNano2012, KonigWeigAPL2012}, a paradigmatic nanoelectromechanical system driven by the interplay of mechanical motion and sequential electron tunneling \cite{GorelikEtAlPRL1998, WeissZwergerEPL1999, BoeseSchoellerEPL2001, ArmourMacKinnonPRB2002, NordEtAlPRB2002, MccarthyEtAlPRB2003, NovotnyEtAlPRL2003, NovotnyEtAlPRL2004, UtamiEtAlPRB2006, NoceraEtAlPRB2011, PradePlateroPRB2012, TonekaboniEtAlArXiv2018, WaechtlerEtAlNJP2019, WaechtlerPRAppl2019, StrasbergPRL2021} (for reviews see Refs.~\cite{ShekhterEtAlJoP2003, GalperinEtAlJoP2007, LaiEtAlFoP2015}). In this work, we analyze this autonomous system and propose a design principle to turn it into a reliable SES. To this end, we characterize the regularity of electron emission through the probability distribution of times between tunneling events. Also known as waiting time distribution (WTD) \cite{BrandesAnn2009, AlbertEtAlPRL2011}, it is a common approach for conventional SES \cite{AlbertEtAlPRL2011, potanina2017electron} and we adapt this tool for use in the context of self-oscillatory systems. For experimentally realistic parameters of existing electron shuttles, we show that the regularity of the SES increases as the coupling to the leads decreases, which is in stark opposition to conventional electron pumps or turnstiles \cite{potanina2017electron, brange2021controlled}. While for these parameters the precision remains rather low, we propose an alternative design  of the tunneling-rate dependency on the shuttle position, which significantly reduces the relative error. Together with its self-regulated character, this system is expected to be less prone to control errors as opposed to existing, non-autonomous SES implementation. 

\section{Mechanism and operating regimes}

The electron shuttle consists of an oscillating quantum dot (QD) located between two fermionic leads, a source and a drain, such that electrons can tunnel between the leads and the QD as depicted in Fig.~\ref{fig:Model}(a). We assume that the QD can accept only a single excess electron
(strict Coulomb blockade regime), which is justified for low temperatures, $k_\mathrm B T \ll E_C$ ($k_\mathrm B$ being the Boltzmann constant and $E_C$ the charging energy), and voltages $|V| \ll E_C/\mathrm{e}$ ($\mathrm e$ being the elementary charge). In the regime $k_\mathrm B T \ll |\mathrm{e}V|$, electron tunneling may be assumed to be unidirectional from source to drain. Experimentally, shuttles have been realized e.g. through nanopillars with a metallic top layer~\cite{KimNJP2010, KimPRL2010, KonigWeigAPL2012, KimEtAlNano2012}.

 The autonomous oscillations of the shuttle arise due to proximity effects of electron tunneling in combination with an electrostatic field generated by the bias voltage $V$ between the leads: As the shuttle approaches the source, the probability of electron tunneling from source to QD exponentially increases. After the electron has tunneled, the charged QD experiences an  electrostatic force pushing it towards the drain, where the electron tunnels from the QD to the lead. Finally, the restoring force of the oscillator pulls the shuttle back to the source and the cycle starts all over. The mechanical friction $\gamma$ compensates the tendency of the shuttle to gradually increase its range of oscillation with each cycle. From the point of view of the drain lead, an electron periodically tunnels into it, in a behavior identical to that of a common SES.

As mentioned previously, we characterize the SES in terms of WTDs, which should peak close to the oscillator period $2\pi/\omega_0\approx 78.5$~ns corresponding to a highly regular emission of electrons separated in time by the shuttling period. Because of the unidirectional tunneling through the system, monitoring the current through a nearby quantum point contact $I_\mathrm{QPC}$  \cite{SchallerKiesslichBrandesPRB2010, wagner2017strong} as depicted in Fig.~\ref{fig:Model}(a), allows to access the waiting time $\tau$ between two consecutive electron jumps into the drain. These jumps can be simulated stochastically \cite{SM} and an example of such a time trace is shown in Fig.~\ref{fig:Model}(b). The WTD can then be constructed, which allows us to distinguish three regimes with typical WTDs shown in Fig.~\ref{fig:Model}(c):

\textit{Transistor regime (I):} For a large tunneling rate $\Gamma$ (magenta solid) the WTD is peaked at a time significantly below the natural oscillation period of the shuttle,  characteristic for a single electron transistor (SET) \cite{davies1992classical, BrandesAnn2009}. 

\textit{Crossover regime (II):} As $\Gamma$ is reduced (brown dashed-dotted), two additional peaks appear located approximately at the oscillation period $2\pi/\omega_0$ and half of it. 

\textit{Shuttle regime (III):} For low $\Gamma$ (blue dashed), a single prominent maximum exists at the oscillation period, characteristic for SES. 

While the transistor and shuttle regime have been identified in previous studies on the electron shuttle \cite{GorelikEtAlPRL1998, NovotnyEtAlPRL2003, NovotnyEtAlPRL2004, WaechtlerEtAlNJP2019}, the crossover regime would have not surfaced  within the mean field description from analysis of the time-averaged electron currents, as the shuttle motion undergoes a Hopf bifurcation from static to shuttling motion \cite{WaechtlerEtAlNJP2019}. The identification of the three regimes requires the use of WTDs as discussed in more detail below.

 \begin{figure*}
\begin{center}
\includegraphics[width=\linewidth]{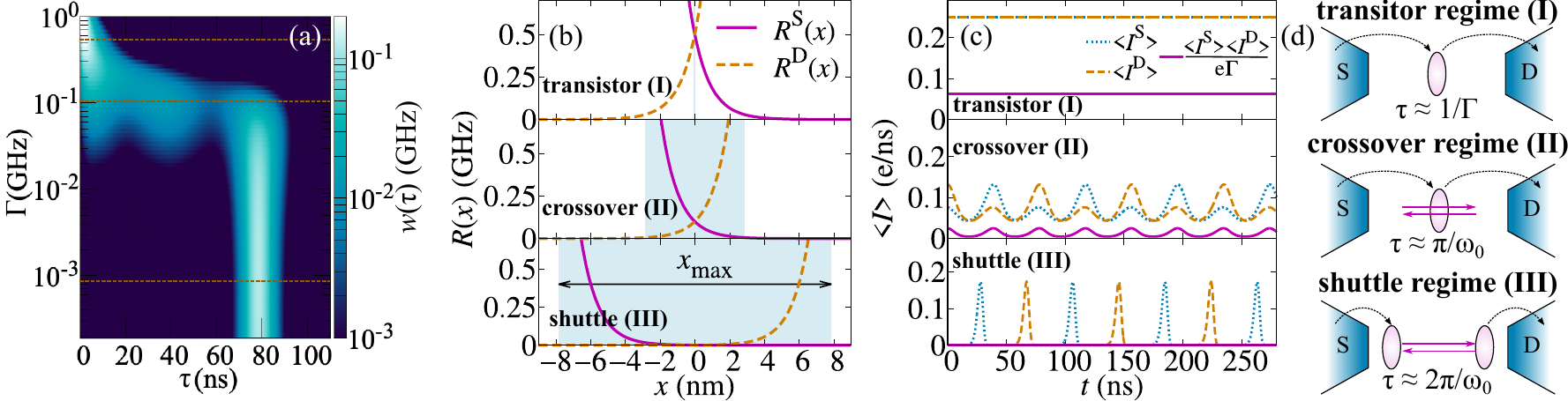}
\end{center}
\caption{(a) Waiting time distribution $w(\tau)$ for $V=70$~mV showing the transition from transistor to shuttle regime as $\Gamma$ is reduced. (b) Tunneling rates from the source (magenta solid) and to the drain (brown dashed) as function of the shuttle position for the parameters marked by the dashed brown lines in panel (a). The blue shaded areas indicate the maximum oscillation amplitude $x_{\mathrm{max}}$ at steady state. (c) Electron current from the source to the QD $\braket{I^\mathrm S}$ (blue dotted), from the QD to the drain $\braket{I^\mathrm D}$ (brown dashed) and the product of both (solid magenta) as function of time for the parameters marked in panel (a). (d) Schematic representation of the interplay of shuttle dynamics and waiting times for the  regimes (I)--(III) as $\Gamma$ is decreased from top to bottom.}
\label{fig:Fig2}
\end{figure*}

\section{Numerical procedure}

We consider here a classical description of the shuttle dynamics, which can be derived from a Markovian quantum master equation \cite{SM}. This classical mean field description also approximates a fully stochastic description very well for the parameters used \cite{SM}. The coupled nonlinear differential equations governing the position $x$ of the shuttle and its charge occupation probabilities $\mathbf p=(p_0,p_1)$ are given by
\begin{equation}
\label{eq:MFEquations}
\ddot x = -\omega_0^2 x-\frac{\gamma}{m} \dot x + \frac{V \mathrm e}{d m} p_1,\quad  \dot{\mathbf{p}} = \left[\begin{matrix}
-R^\mathrm{S}(x) & R^\mathrm{D}(x) \\
R^\mathrm{S}(x) & -R^\mathrm{D}(x)
\end{matrix}\right] \mathbf p, 
\end{equation}
where $\omega_0$ and $m$ are the oscillator frequency and mass, respectively. The electrostatic field an electron experiences may be approximated by the field of a plate capacitor $V/d$ with distance $d$ between the leads \cite{GorelikEtAlPRL1998, WaechtlerEtAlNJP2019}. While the equation of motion of the oscillator is linear, the dynamics of $\mathbf{p}$ depend exponentially on the position of the shuttle, i.e., $R^\mathrm{S/D}(x) = \Gamma \exp(\mp x/\lambda)$ with the bare tunneling rate $\Gamma$ and a characteristic tunneling length $\lambda$. This nonlinear dependency is necessary for the electron shuttle to reproduce self-oscillations, and thus operate autonomously.

While for time-independent systems the WTD can be obtained straightforwardly \cite{BrandesAnn2009}, self-oscillatory systems demand more careful considerations. For the case at hand, the difficulty arises as  the occupation probability $p_1$ is dynamically coupled to the shuttle position [cf. Eq.~(\ref{eq:MFEquations})]. However, in the long-time limit, the system may reach a (periodic) steady state which allows us to define a WTD. To determine the probability density $w(\tau)$ of a waiting time $\tau$ between two consecutive tunneling events into the drain, we require the  electron current from the QD to the drain at time $t'$, $\langle I^\text D(t')\rangle/ \mathrm e  =R^\mathrm{D}[x(t')] p_1(t')$, and the conditional current at time $t'+\tau$, $\langle I^\text D(\tau|t')\rangle/ \mathrm e  = R^\mathrm{D}[x(\tau|t')] p_1(\tau|t')$.  The occupation probability $p_1(\tau|t')$ conditioned on an electron having tunneled into the drain for the last time at $t'$ is determined by solving Eq.~(\ref{eq:MFEquations}) for the initial conditions $\mathbf p(0|t')=(1,0)$, $x(0|t')=x(t')$ and $\dot x(0|t')=\dot x(t')$ with a modified rate matrix, where the entry $R^\mathrm D(x) \equiv 0$ [while keeping the entry $-R^\mathrm D(x)$]. Then, the WTD is defined as 
\begin{equation}
\label{eq:WTD}
w(\tau) = \frac{\mathcal N }{T_\mathrm{osc}{\mathrm e}^2}\int\limits_{t_0}^{t_0+T_\mathrm{osc}} \left<I^\text D(t')\right> \left<I^\text D(\tau|t')\right> dt', 
\end{equation}
where $\mathcal N$ is a normalization constant and $t_0$ is a random starting time within the periodic steady state with period $T_\mathrm{osc}$. Note that in general the shuttling period $T_\mathrm{osc}$ slightly differs from the uncoupled oscillator period $2\pi/\omega_0$. The WTD (\ref{eq:WTD}) may equivalently be defined in terms of the current from the source to the shuttle $\langle I^\text S(t')\rangle/ \mathrm e  =R^\mathrm{S}[x(t')] p_0(t')$.

In this work, we use parameters estimated from experimental realizations of the electron shuttles. Throughout, we use $\omega_0 = 80$~MHz \cite{KimPRL2010}, $m=20\times 10^{-19}$~kg  \cite{KimPRL2010, KimEtAlNano2012}, $\lambda =1$~nm \cite{KimEtAlNano2012}, $d=100$~nm \cite{KimAPL2007, KimNJP2010} and $\gamma = 10^{-13}$ kg/s.

\section{Results and discussion}
\subsection{Characterization in terms of WTDs}

We start our analysis by investigating the influence of the bare tunneling rate $\Gamma$ on the WTD. As shown below, the qualitative behavior of the system is not strongly affected by the choice of the applied voltage $V$ as long as unidirectionality is preserved and we choose for now a typical value for experiments of $V=70$~mV \cite{KimPRL2010}. In Fig.~\ref{fig:Fig2}(a) we show $w(\tau)$ for a large range of $\Gamma$ (note the logarithmic scale). Here, we observe a smooth transition between the regimes described above: the transistor regime (I) at large $\Gamma$ with a WTD peaked at small $\tau$, the crossover regime (II) with broad WTD and triple-peak structure, and the shuttle regime (III) at small $\Gamma$ with a narrow WTD peaked at $T_\mathrm{osc}$. 

The shape of the WTD of the three different regimes is strongly correlated with the oscillation amplitude $x_\mathrm{max}$ of the shuttle and the overlap of the tunneling rates. To illustrate this, we show the tunneling rates $R^\mathrm{S/D}(x)$ in Fig.~\ref{fig:Fig2}(b) for three different values of $\Gamma$ marked by the brown dashed lines in panel (a) together with the range covered by the shuttle oscillations at steady state (blue shaded area). Additionally, we plot in Fig.~\ref{fig:Fig2}(c) the electron currents $\braket{I^\mathrm S}$ and $\braket{I^\mathrm D}$, and their product $\braket{I^\mathrm S}\braket{I^\mathrm D}$ as function of time. The latter quantifies the ability of electrons to immediately tunnel out of the QD after having tunneled in (note the relationship to the WTD). Interestingly, this quantity is proportional to the variance of the charge occupation due to the symmetry of the tunneling rates, i.e., $\braket{I^\mathrm S}\braket{I^\mathrm D}/\mathrm e^2 = \Gamma^2 (\braket{q^2}-\braket{q}^2)$. 

In the transistor regime (I) shown in the top panels ($\Gamma = 1.0$ GHz) the QD does not oscillate [cf. top panel of Fig.~\ref{fig:Fig2}(d)]. Consequently, the currents  from source (blue dotted) and to drain (brown dashed) shown in panel (c) are constant and the WTD is equivalent to the one we expect from a SET given by $w_\mathrm{SET}(\tau) = \Gamma^2 t \exp(-\Gamma t)$ \cite{davies1992classical, BrandesAnn2009}, see solid magenta line in Fig.~\ref{fig:Model}(c). 

On the other hand, in the shuttle regime (III) shown in the bottom panels of Fig.~\ref{fig:Fig2}(b)--(d) for $\Gamma = 10^{-3}$~GHz, the QD oscillates back and forth between the two leads. Fig.~\ref{fig:Fig2}(b) shows that there is no considerable overlap of the tunneling rates in this regime. Thus, tunneling occurs in the proximity of the respective lead, as indicated by the periodic peaks of the electron currents [cf. Fig.~\ref{fig:Fig2}(c)]. As virtually all electrons are shuttled, the quantity $\braket{I^\mathrm S}\braket{I^\mathrm D}\simeq0$ practically vanishes. A schematic representation is shown in the bottom panel of Fig.~\ref{fig:Model}(d). The resulting WTD is peaked at the shuttling period indicating the SES character. 

In the crossover regime (II), the WTD exhibits features from both regimes (I) and (III): A peak at small $\tau$ from electron tunneling and a peak at large $\tau$ from electron shuttling. In addition, this regime exhibits a feature that is not present in either of the previous regimes: During one oscillation period the QD is twice tunnel coupled to both leads simultaneously, once during the forward motion and once during the backward motion [see middle panel of Fig.~\ref{fig:Fig2}(b) for $\Gamma = 0.1$~GHz]. Thus, two consecutive tunneling events into the drain may be separated by half the oscillation period resulting in a peak of the WTD at $\tau\approx \pi/\omega_0$. All three features of the WTD are also clearly observed in the structure of the electron currents from and to the leads: Both currents $\braket{I^\mathrm S}$ and $\braket{I^\mathrm D}$ remain finite at all times (transistor regime) and the large peaks of the currents are separated by the oscillation period (shuttle regime). Additionally, small peaks appear at all half periods. Hence, $\braket{I^\mathrm S}\braket{I^\mathrm D}$ (magenta solid) peaks at half the  period.  Notably, the crossover regime described in this work differs substantially from a crossover regime found for a quantum mechanical description of the electron shuttle \cite{NovotnyEtAlPRL2003, NovotnyEtAlPRL2004}, where it refers to the coexistence of oscillatory and non-oscillatory shuttle motion.

\begin{figure}
\begin{center}
\includegraphics[width=\columnwidth]{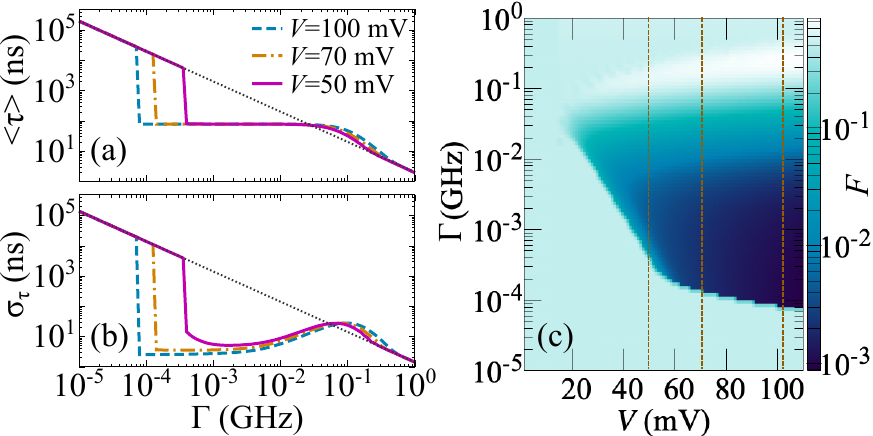}
\end{center}
\caption{(a) Averaged waiting time $\braket{\tau}$ and (b) its standard deviation $\sigma_\tau$ as function of the bare tunneling rate $\Gamma$ for three different values of the bias voltage. The gray dotted lines correspond to the respective values of an SET. (c) Density plot of the Fano factor $F$ as function of $V$ and $\Gamma$ demonstrating that the shuttle performs as SES over a wide parameter range. The parameters marked by the dashed brown lines correspond to panels (a) and (b).}
\label{fig:Fig3}
\end{figure}

\subsection{Fano factor analysis}

The previously discussed shape of the WTDs directly translates to the statistical properties of the waiting time $\tau$ in terms of its average $\braket{\tau}=\int_0^\infty \tau w(\tau) d\tau$ and its standard deviation $\sigma_\tau =\sqrt{\langle (\tau-\langle \tau\rangle)^2\rangle}$. For three different applied bias voltages $V$, we show these two quantities as a function of $\Gamma$ in Fig.~\ref{fig:Fig3}(a) and (b), respectively, together with the values expected for an SET as gray dotted lines. The mean waiting time $\braket{\tau}$ clearly reflects the three regimes as a function of $\Gamma$: Regime (I) for large $\Gamma$ with $\braket{\tau_\mathrm{SET}}=2/\Gamma$ (which does not coincide with the peak of the WTD located at $1/\Gamma$  due  to the asymmetry of the distribution) independent of the bias voltage, regime (II) where $\braket{\tau}$ deviates from SET value, and regime (III) where the average waiting time takes on a constant value of $\braket{\tau}=2\pi/\omega_0$ independent of $\Gamma$. Interestingly, as $\Gamma$ is further reduced, $\braket{\tau}$ increases sharply. Here, the system undergoes a transition back to the transistor regime. Thus, the WTD is peaked again at $1/\Gamma$, which is this time significantly larger than the shuttling period. 

While the average waiting time $\braket{\tau}$ is a useful tool to explore the operating regimes of the system, it is not sufficient to assess its performance as a SES. For this, we turn to $\sigma_\tau$ shown in Fig.~\ref{fig:Fig3}(b), where the black dotted line corresponds to the SET value of $\sigma_\mathrm{SET}=\sqrt{2}/\Gamma$. Within the shuttle regime (III), electron emission becomes more deterministic as $\Gamma$ is reduced until the sharp breakdown of shuttling. A common figure of merit to determine the regularity of a SES is the Fano factor $F=\sigma_\tau^2/\braket{\tau}^2$ \cite{davies1992classical, BrandesAnn2009}. As shown in Fig.~\ref{fig:Fig3}(c), $F$ is significantly smaller than the value for an SET ($F=1/2$) over a wide range of $\Gamma$ and $V$, and exhibits its smallest value at the onset of oscillations (small $\Gamma$). 

\subsection{Dependency on friction and asymmetry \label{IIIC}}

\begin{figure}[h]
\begin{center}
\includegraphics[width=\columnwidth]{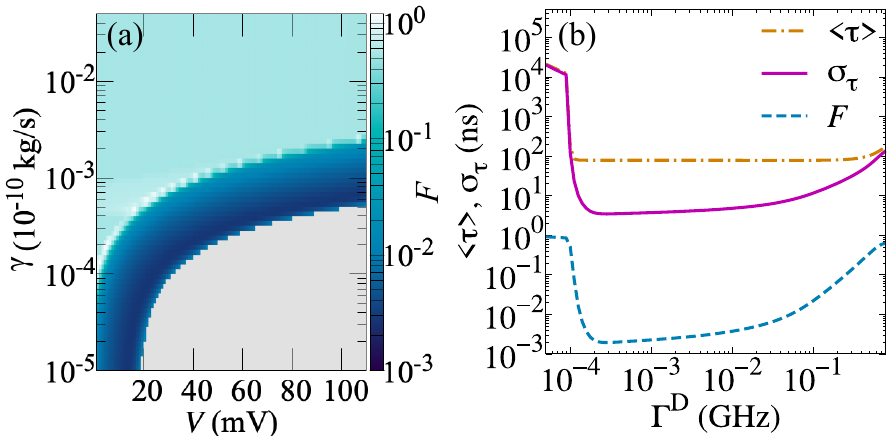}
\end{center}
\caption{(a) Fano factor as function of the bias voltage $V$ and the friction $\gamma$ for $\Gamma = 10^{-3}$~GHz. (b) Influence of different tunneling rates between source and drain on the average waiting time $\braket{\tau}$, its standard deviation $\sigma_\tau$, and the Fano factor $F$ for $\gamma = 10^{-13}$~kg/s and $V = 70$~mV.}
\label{fig:gamma}
\end{figure}

Beyond $\Gamma$ and $V$, which can be easily adjusted in experiments by means of gate voltages, the WTD may also sensitively depend on fabrication parameters such as the friction $\gamma$, directly related to the quality factor of the oscillator. In Fig.~\ref{fig:gamma}(a) we show $F$ as a function of $\gamma$ and $V$ for a fixed value of $\Gamma=10^{-3}$~GHz. If the friction $\gamma$ is too large the shuttle oscillations are damped and the system remains equivalent to an SET. As $\gamma$ is reduced, a sudden transition to the shuttle regime (III) is observed. Within this regime, increasing the voltage $V$ reduces $F$ and thus increases the performance as SES. In the gray area the amplitude of oscillations become so large ($x_\mathrm{max}>15$~nm) that the effective tunneling rates $R^\mathrm{S/D}(x)$ are in the order of $10^4$~GHz and our weak coupling description becomes invalid. For comparison, experimental realizations report of $x_\mathrm{max}=5-8$~nm \cite{KimEtAlNano2012}. However, we observe that over a broad range of the friction, the shuttle regime with narrow WTD can be achieved by adjusting the bias voltage $V$.

Additionally, the question arises whether electron emission can become more deterministic through an asymmetry in the design. For simplicity we have assumed equal tunneling rates for the source and drain, i.e., $\Gamma^\mathrm{S/D}\equiv \Gamma$. However, an unbalanced situation may result in a narrower WTD. In Fig.~\ref{fig:gamma}(b) we show the dependency of $\braket{\tau}$, $\sigma_\tau$ and $F$ as a function of the tunneling rate to the drain $\Gamma^\mathrm D$ for a fixed value of the tunneling rate from the source of $\Gamma^\mathrm S=10^{-3}$~GHz. While $\sigma_\tau$ and thus $F$ may be further decreased by adjusting $\Gamma^\mathrm D$, its effect in the shuttle regime remains rather small.

\section{Towards high-accuracy single-electron sources}

\begin{figure}[t]
\begin{center}
\includegraphics[width=\columnwidth]{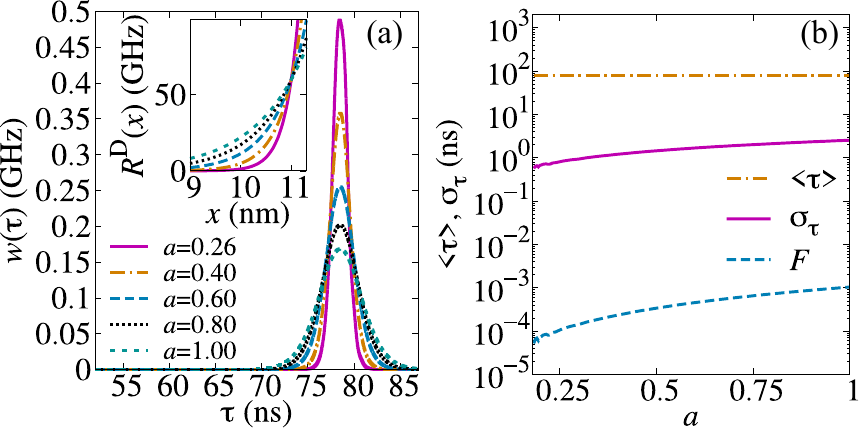}
\end{center}
\caption{(a) Waiting time distribution $w(\tau)$ for different values of $a$ for a fixed value of $x_d=11.0$~nm. Inset: Dependency of the tunneling rates on the shape parameter $a$. (b) Average waiting time $\braket{\tau}$, the standard deviation $\sigma_\tau$, and the Fano factor $F$ as function of $a$ for a fixed value of $x_d=11.0$~nm. Other parameters: $\Gamma = 10^{-4}$~GHz and $V = 100$~mV.}
\label{fig:aParam}
\end{figure}

For the experimentally realistic parameters used in this work, the precision of the SES measured in terms of the Fano factor remains in the regime of $F\approx 10^{-3}$. However, for metrology purposes smaller relative errors of the charge transmission are needed \cite{pekola2013single}. While the adjustment of fabrication parameters such as the friction $\gamma$ or an asymmetric design of the bare tunneling rates $\Gamma$ only slightly improves the Fano factor (see Sec.~\ref{IIIC}), the overall shape of the rates $R^\mathrm{S/D}(x)$ has a much greater impact. In order to sharpen the WTD, the rates should remain small while the QD moves from one lead to the other and increase rapidly close to the maximum oscillation amplitude $x_\mathrm{max}$. To achieve such a behavior, we parametrize the rates via 
\begin{equation}
R^\mathrm{S/D}(a, x) = \Gamma \exp\left\{\left[\mp x + x_d  (a-1)\right]/(a \lambda)\right\},
\end{equation}
such that $a$ controls their slope at $x_d$ (see inset of Fig.~\ref{fig:aParam}). For $a=1$ we recover the original tunneling rates.

In Fig.~\ref{fig:aParam} (a) we show the WTDs for different values of $a$ for a bias voltage of $V=100$~mV and $\Gamma = 10^{-3}$~GHz. Here, we choose $x_d = 11.0$~nm slightly below $x_\mathrm{max} = 11. 2$~nm, which we have found to have the greatest impact. As $a$ decreases the WTD becomes narrower while the peak of the distribution remains at the same position, which is also confirmed by the average waiting time $\braket{\tau}$ (orange dash-dotted) and the standard deviation $\sigma_\tau$ (solid magenta) shown in panel (b). As expected the Fano factor (blue dashed) also significantly decreases, here more than one order of magnitude. Adjusting the shape of the electron tunneling rates thus seems like a promising route towards a high precision SES based on self-oscillation. 

A redefinition of our parametrization may be helpful to better illustrate the experimental implications of our finding. By noticing that we may rewrite $R^\mathrm{S/D}(a, x)$ as $\Gamma'(a) \exp{[\mp x/(a\lambda)]}$
it is clear that parameter $a$ affects both the tunneling length $\lambda$ and the bare tunneling rates $\Gamma'$. Our conclusion that a small $a$ would improve the SES function translates into a design that should aim at short tunneling lengths while accordingly adjusting transmission coefficients. The parameter $a$ is in this sense a shorthand notation to facilitate analysis, but both properties should be optimized in experimental realizations.

The form of the rates is consistent with the assumption of a simple picture of electrons tunneling through a thin insulating barrier. This may be modeled by a single-particle rectangular potential whose height is determined by the work function of the material, i.e. the energy required to extract an electron from it. Under this picture, the property that most strongly affects the tunneling length $\lambda$ is indeed the work function. Based on typical values, the choice of the material alone may provide a factor of $a=1/3$. Doping, surface-coating of the leads with thin layers of insulating material or control through external electric fields introduce additional means to reduce $a$.

To achieve the accordance of the bare tunneling rates $\Gamma'$, in addition to the work function one may take advantage of the distance to the leads and even of their shape. A more accurate analysis, for instance by means of Bardeen's formalism for tunneling, could be sufficient to provide experimentalists with specific designs that would adjust these values appropriately.

A complementary route might be to consider the non-Markovian regime by assuming highly peaked spectral densities, which may assist the shuttle to become more deterministic \cite{StrasbergEtAlNJP2016, StrasbergEtAlPRB2018, SchallerEtAlPRB2018, NazirSchallerBook2018, RestrepoPRB2019, gurvitz2019generalized}. Exploring this regime as well as the specific design of experimental realizations is subject to further work on autonomous SES. 

\section{Conclusions}We have studied the capability of a self-oscillatory system to perform as an autonomous SES, where the internal feedback mechanism of electron tunneling and mechanical motion removes the necessity to periodically vary the tunneling barriers externally. Over a wide range of parameters we have shown that the electron shuttle exhibits a narrow WTD. Closely modeling the parametric regime of already existing shuttles, the best performance is achieved right above the onset of self-oscillation. To significantly increase the regularity of electron transmission, we find that a change in the overall shape of the tunneling rates is sufficient, which only involves accurate location of the leads with respect to the shuttle and careful adjustment of its tunneling length. Additionally, the WTD reveals information about the system dynamics inaccessible through the time-averaged current, specifically, a smooth transition from a transistor to a shuttle regime separated by a complex crossover. 

\begin{acknowledgments}
We thank M.T. Eiles for carefully reading the paper. C.W.W. acknowledges support from the Max-Planck Gesellschaft via the MPI-PKS Next Step fellowship. J.C. acknowledges support from Ministerio de Ciencia, Innovación y Universidades (Spain) (“Beatriz Galindo” Fellowship BEAGAL18/00078).
\end{acknowledgments}


%

\newpage
\onecolumngrid
\section{SUPPLEMENTAL MATERIAL}

In this supplemental material we provide (A) the quantum master equation and the derivation of the mean field equations, and (B) the classical stochastic model and its comparison to the mean field model.

\section{(A) Derivation of the mean field equations from a generalized quantum master equation}
Our starting point is the full quantum master equation for the combined oscillator-dot system as derived in Ref.~\cite{NovotnyEtAlPRL2003}:
\begin{align}
\label{eq:QuantumME}
\dot{\hat\varrho} (t) &= -\frac{\mathrm{i}}{\hbar}\left[\frac{\hat p^2}{2m}+ \frac{1}{2} m\w^2\hat x^2 + \left(\epsilon -\frac{\mathrm e V \hat x}{d}\right) \hat c^\dagger \hat c,\hat \varrho(t)\right] + \left(\mathcal L_\mathrm{drive}+\mathcal{L}_\mathrm{damp}\right)\hat \varrho(t),\\
\mathcal L_\mathrm{drive}\hat \varrho &= -\frac{\Gamma}{2}\left(\hat c \hat c^\dagger e^{-\hat x/\lambda}\hat \varrho -2\hat c^\dagger  e^{-\hat x/2\lambda}\hat \varrho e^{-\hat x/2\lambda}\hat c +\hat \varrho e^{-\hat x/\lambda}\hat c\hat c^\dagger + \hat c^\dagger \hat c e^{\hat x/\lambda}\varrho -2 \hat c e^{\hat x/2\lambda}\hat \varrho e^{\hat x/2\lambda}\hat c^\dagger +\hat \varrho e^{\hat x/\lambda}\hat c^\dagger \hat c \right),\\
\mathcal L_\mathrm{damp}\hat \varrho &= -\mathrm i\frac{\gamma}{2m\hbar}\left[\hat x,\left\{\hat p,\hat \varrho\right\}\right]-\frac{\gamma \omega_0}{\hbar}\left(\bar N + \frac{1}{2}\right)\left[\hat x,\left[\hat x,\hat \varrho\right]\right].
\end{align}
The operator $\hat c^\dagger$ ($\hat c$) creates (annihilates) an electron on the dot and we explicitly use a hat notation to denote operators. Furthermore, $\bar N=(e^{\beta \hbar \omega}-1)^{-1}$ is the Bose-Einstein distribution. We note that the master equation (\ref{eq:QuantumME}) is derived in the `high-bias' limit, such that the Fermi functions of the source and drain may be replaced by $1$ and $0$, respectively, which we also assume in this work.

In order to arrive at the mean field equations of the main manuscript, we approximate $\exp(\pm \hat x/\lambda)\approx \exp(\pm \braket{\hat x}/\lambda)$ \cite{WaechtlerEtAlNJP2019}, such that $\mathcal L_\mathrm{drive}\hat \varrho$ in Eq.~(\ref{eq:QuantumME}) only depends on the average position $\braket{x}$ of the oscillator. In this way, we are able to derive a closed set of equations for the average position, momentum and dot occupation of the system:
\begin{equation}
\begin{aligned}
\frac{d}{dt}\braket{x} &= \braket{p}/m,\\
\frac{d}{dt}\braket{p} &= -m\omega_0^2 \braket{x}-\frac{\gamma}{m}\braket{p} + \frac{\mathrm e V}{d}p_1,\\
\frac{d}{dt}\mathbf{p}&=\Gamma \left(\begin{matrix}
-e^{-\braket{x}/\lambda} &  e^{\braket{x}/\lambda} \\
e^{-\braket{x}/\lambda} & - e^{\braket{x}/\lambda}
\end{matrix}\right)\mathbf p,
\end{aligned}
\end{equation}
where $\mathbf{p}=(\braket{cc^\dagger},\braket{c^\dagger c})$. These equations correspond to the set of coupled nonlinear differential equations of the main manuscript. 

\section{(B) Classic stochastic equations}

From the full quantum master Eq.~(\ref{eq:QuantumME}) of the previous Sec.~(A), one can derive a classical coupled Fokker-Planck and master equation by taking the limit $\hbar \to 0$ as shown in the supplemental material of Ref.~\cite{StrasbergPRL2021}. In this classical limit, the dynamics of the probability density $P_q(x,v;t)$ to find the shuttle at position $x$ with velocity $v$ and charge $q$ is given by
\begin{equation}
\label{eq:FPE}
\frac{\partial P_q(x,v;t)}{\partial t}= \left\{-v\frac{\partial}{\partial x}+\frac{\partial }{\partial v}\left[\omega_0^2 x + \frac{\gamma}{m}v-\frac{\mathrm e V}{d m}q+\frac{\gamma}{\beta m^2}\frac{\partial}{\partial v}\right]\right\}P_q(x,v;t) + \sum\limits_{q'}R_{qq'}(x) P_{q'}(x,v;t).
\end{equation}

The tunneling rates $R_{qq'}(x)$ are given by the $R_{10}(x) = - R_{00}(x) =  R^\mathrm S(x)$ and $R_{01}(x)= -R_{11}(x) = R^\mathrm D(x)$. The dynamics at the trajectory level might equivalently be represented by the stochastic differential equations \cite{WaechtlerEtAlNJP2019}
\begin{align}
dx &= vdt,\label{eq:Langevin1}\\
dv &= \left(-\omega_0^2 x - \frac{\gamma}{m}v + \frac{V\mathrm e}{dm}q\right)dt + \sqrt{\frac{2\gamma}{\beta m^2}} dB(t), \label{eq:Langevin2}\\
dq &= \sum\limits_{q'}(q'-q)dN_{q'q}(x,t).\label{eq:Langevin3}
\end{align}
Here, $dB(t)$ represents a Wiener process with zero mean $\mathbb E[dB(t)]=0$ and variance $\mathbb E[dB(t)^2]=dt$, where $\mathbb E[\bullet]$ denotes an average of the stochastic process. Furthermore, the stochastic electron tunneling is described by the independent Poisson increments $dN_{q'q}(x,t)\in \{0,1\}$ obeying the statistics $\mathbb E[dN_{q'q}(x,t)]=R_{q'q}(x)dt$ and $dN_{q'q}(x,t)dN_{\tilde qq}(x,t) = \delta_{q'\tilde q}dN_{q'q}(x,t)$. For a detailed discussion on the equivalence of Eqs.~(\ref{eq:FPE}) and (\ref{eq:Langevin1})-(\ref{eq:Langevin3}) we refer to Ref.~\cite{WaechtlerEtAlNJP2019}. We use the stochastic differential Eqs.~(\ref{eq:Langevin1})-(\ref{eq:Langevin3}) to simulate the time trace of Fig.~1(b) from the main manuscript.

\begin{figure}[h]
\begin{center}
\includegraphics[width=0.8\columnwidth]{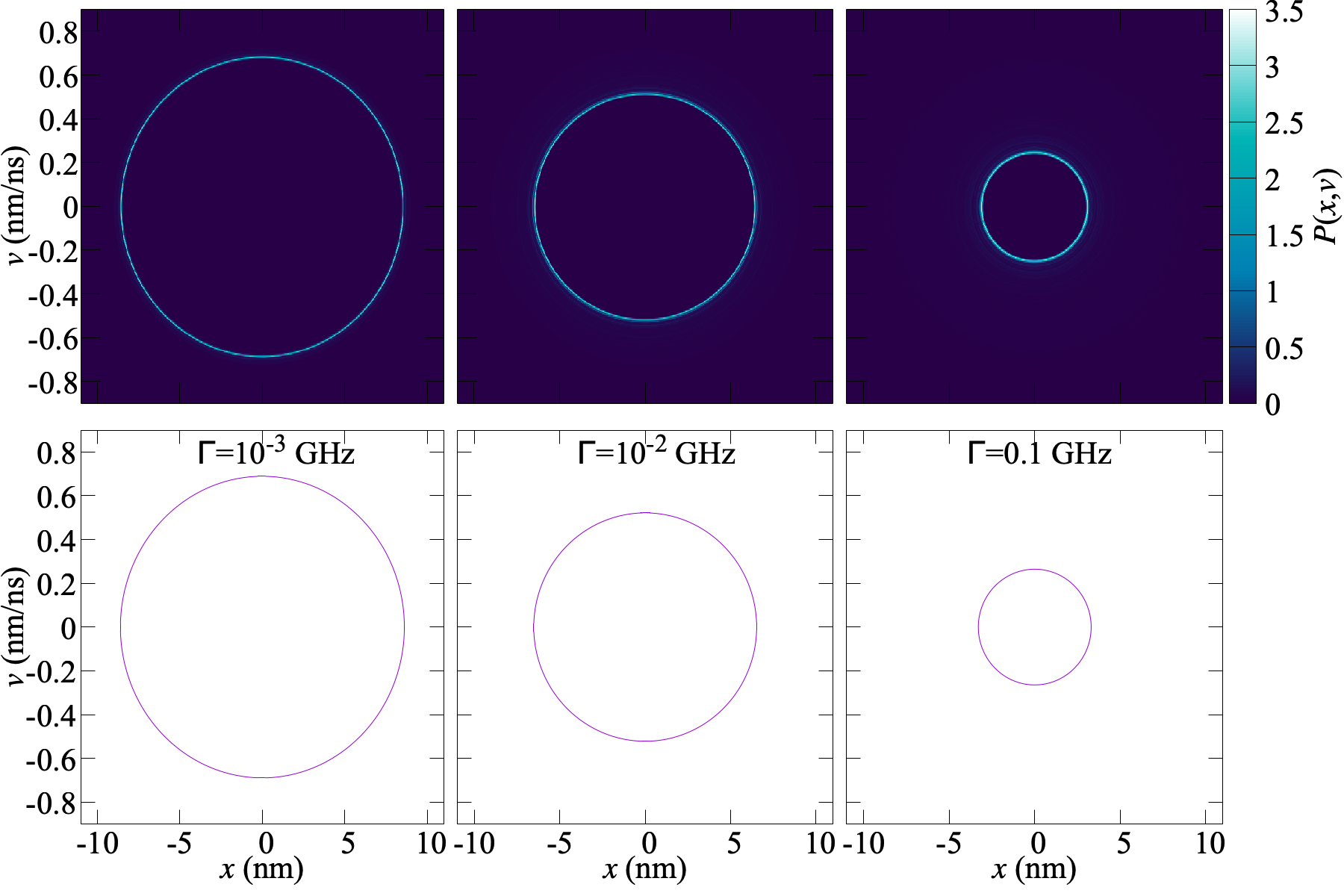}
\end{center}
\caption{Comparison between a full stochastic description (top panels) and the mean field approximation (bottom panels) of the shuttle dynamics showing that the mean field level is a sufficient description for the parameters used in the main manuscript. Here, we used $V=70$ mV.}
\label{fig:Histogramm}
\end{figure}

Similar to Sec. (A), we can derive the mean field equations of the main manuscript by approximating the tunneling rates  $R^\mathrm{S/D}(x)\approx R^\mathrm{S/D}(\braket{x})$. However, the question arises how well the dynamics of the system is approximated for the parameters used. To this end, we compare the stochastic evolution to the one governed by Eq.~(1) of the main manuscript. In Fig.~\ref{fig:Histogramm} we compare the probability density $P(x,v) = \sum_q P_q (x,v)$ at steady state, i.e. for $t\to \infty$, to the phase-space trajectories of the deterministic mean field equations for three different values of the bare tunneling rate $\Gamma$. Due to the low temperature (here we used $T=1$ K), the probability density $P(x,v)$ remains a sharp ring, well represented by the deterministic description, and thermal fluctuations may be neglected. 

\end{document}